\documentclass[10pt,twocolumn,english,twocolumn,aps,prd,longbibliography]{revtex4-1}
\usepackage[T1]{fontenc}
\usepackage[latin9]{inputenc}
\usepackage{float}
\usepackage{textcomp}
\usepackage{amsmath}
\usepackage{amssymb}
\usepackage{graphicx}
\usepackage{esint}

\makeatletter
\usepackage{lmodern}
\usepackage[T1]{fontenc}
\usepackage{prettyref}
\usepackage[unicode=true,
 bookmarks=false,
 breaklinks=false,pdfborder={0 0 1},colorlinks=false,colorlinks=true,
  urlcolor=blue,
  linkcolor=blue,
  citecolor=blue]{hyperref}
\usepackage{textgreek}
\usepackage{inputenc}
\usepackage{slashed}

\makeatother

\usepackage{babel}
\begin{document}
\title{Investigation of two photon emission in strong field QED using channeling
in a crystal}
\author{Tobias N. Wistisen}
\affiliation{Max-Planck-Institut f{\"u}r Kernphysik, Saupfercheckweg 1, D-69117, Germany}
\begin{abstract}
We investigate the 2nd order process of two photons being emitted
by a high-energy electron dressed in the strong background electric
field found between the planes in a crystal. The strong crystalline
field combined with ultra relativistic electrons is one of very few
cases where the Schwinger field can be experimentally achieved in
the electron's rest frame. The radiation being emitted, the so-called
channeling radiation, is a well studied phenomenon. However only the
first order diagram corresponding to emission of a single photon has
been studied so far. We elaborate on how the 2 photon emission process
should be understood in terms of a two-step versus a one-step process,
i.e., if one can consider one photon being emitted after the other,
or if there is also a contribution where the two photons are emitted
'simultaneously'. From the calculated full probability we see that
the two-step contribution is simply the product of probabilities for
single photon emission while the additional one-step terms are, mainly,
interferences due to several possible intermediate virtual states.
These terms can contribute significantly when the crystal is thin.
Therefore, in addition, we see how one can, for a thick crystal, calculate
multiple photon emissions quickly by neglecting the one-step terms,
which represents a solution of the problem of quantum radiation reaction
in a crystal beyond the usually applied constant field approximation.
We explicitly calculate an example of 180 GeV electrons in a thin
Silicon crystal and argue why it is, for experimental reasons, more
feasible to see the one-step contribution in a crystal experiment
than in a laser experiment.
\end{abstract}
\maketitle
Strong field QED is the study of physical processes that take place
in a strong background field and nonlinear effects of quantum nature
arise when the size of the Lorentz invariant parameter 
\begin{equation}
\chi=e\sqrt{(F_{\mu\nu}p^{\nu})^{2}}/m^{3},\label{eq:chi}
\end{equation}
is on the order of unity, which is the ratio of the electromagnetic
field experienced in the electron's rest frame compared to the Schwinger
field strength $E_{\text{Sch}}=1.32\times10^{18}\,\text{V/m}$. Here
$e$ is the elementary charge, $m$ the electron mass, $F_{\mu\nu}$
the electromagnetic field tensor of the background field and $p^{\nu}$
the electron 4-momentum. We use natural units such that $\hbar=c=1$,
$\alpha=e^{2}$. Lindhard was one of the first to realize that when
high energy charged particles are aimed close to the direction along
an axis or plane in a crystal, the charged particle can become transversely
trapped \cite{Lind65}. Later it was studied how this motion leads
to radiation emission called channeling radiation, especially relevant
for electrons and positrons. This is well-studied both experimentally
\cite{BAK1985491,BAK1988525,PhysRevLett.43.1723,1402-4896-24-3-015,ANDERSEN1982209,PhysRevB.31.68,PhysRevLett.42.1148,PhysRevD.86.072001,RevModPhys.77.1131}
and theoretically \cite{KUMAKHOV197617,kumakhov1977theory,1402-4896-24-3-015,SAENZ198190,KIMBALL198569}.
Crystal channeling represents one of the only phenomena where the
Schwinger field can be experimentally achieved in the electron's rest
frame \cite{Belk86b,ANDERSEN1982209,Esbe10,wistisen2018experimental},
with the only other example being the famous E-144 SLAC experiment
on non lienar Compton scattering \cite{bula1996observation} using
relativistic electrons colliding with a laser beam. Crystals with
ultra relativistic electrons or positrons therefore present a unique
possibility to study physics in such strong fields. However a calculation
from first principles of emission of more than 1 photon has not been
carried out for crystal channeling. The recent studies of 2 photon
emission in the collision of relativistic electrons with a laser pulse
\cite{PhysRevD.85.101701,PhysRevLett.110.070402,dinu2018single,PhysRevA.91.033415}
show that the emission of 2 photons is not exactly the product of
probabilities for each emission, however under certain conditions
it is an acceptable approximation. The experimental verification of
such results are however complicated in the case of the laser pulse
colliding with an electron bunch because any two (or more) emitted
photons cannot be known to be emitted by the same electron. In crystal
experiments as in e.g. \cite{wistisen2018experimental}, it is standard
that each incoming particle is recorded as a separate event, and therefore
the measured outgoing photons are sure to stem from the single incoming
particle. Therefore, in this paper, we will calculate the emission
of 2 photons during electron channeling in a crystal, which could
potentially be studied experimentally in an experiment similar to
the one seen in \cite{wistisen2018experimental}, however with a modified
setup to allow for the detection of an additional photon. For the
theory of channeling radiation, in particular the development of the
semi-classical operator method by Baier et. al. \cite{baier1968processes}
stands out, and has been extensively applied to the phenomenon of
channeling \cite{Baier1998}. 
\begin{figure}[t]
\includegraphics[width=1\columnwidth]{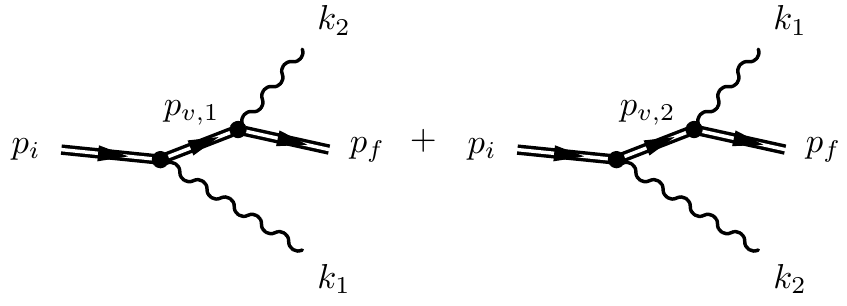}

\caption{The Feynman diagrams corresponding to the process under study. The
double fermion lines correspond to positron solutions of the Dirac
equation in the background field of the inter planar crystal potential.}
\end{figure}
This method allowed to include quantum effects such as the electron
spin and the photon recoil, which are important when $\chi$ is no
longer small, while needing only the classical trajectory of the electron/positron
in the external field. The authors of this method, seeking analytical
results, in most applications to channeling, applied the approximation
of the local constant field which greatly simplifies calculations.
The constant field approximation means that while a particle moves
in an external field, which is not constant, one applies the result
of the constant field formula locally, i.e. in a small time step.
Effectively this means neglecting that the radiation emitted before
or after can interfere with this radiation. This is valid only for
certain parameters of fields and particle energies. However the semi-classical
operator method can be used to calculate the radiation emission under
general circumstances without much effort, also when the constant
field approximation is no longer valid \cite{PhysRevD.90.125008,PhysRevD.92.045045},
which with modern computing power makes it one of the most powerful
methods to calculate the radiation emitted by ultra-relativistic electrons
in a general field configuration. There are caveats however, which
are two-fold. Firstly, the notion of a classical trajectory should
make sense. Or, in other words, the quantum numbers associated with
the motion should be large, a subject recently studied in \cite{Wistisen2019,PhysRevA.98.022131}.
Secondly, the derivation starts out from the first-order diagram of
a dressed electron emitting a single photon. Therefore the emission
rate of two, or more, photons can not be predicted by this method
without approximations. The emission of a single photon yields a rate,
an emission probability per unit time, and as such one can construct
the probability for emitting several photons by applying this rate
for each consecutive emission. In this way, the probability to emit,
e.g., two photons would be proportional to time, or thickness of the
crystal, squared, and so on. We will call this process the 'cascade'
process. Herein lies an approximation, where interference between
different emissions is neglected. We show that the two-photon emission
probability contains the cascade along with one-step terms which scale
linearly with the crystal thickness. Therefore, for sufficiently thin
crystals, these one-step terms will become important. This phenomenon
is also discussed in pair production of electron/positron pairs from
high energy photons in a strong field where one also distinguishes
between the two-step and the one-step, or 'trident' process. This
has been investigated in crystals in \cite{Esbe10} and has received
renewed interest with the prospect of studying such phenomena in high-intensity
laser fields \cite{PhysRevLett.105.080401,PhysRevLett.106.020404,acosta2019laser,PhysRevD.98.016005,PhysRevD.97.036021,PhysRevD.98.116002}.
In this paper we make quantitative calculations of the angularly integrated
probability, differential in photon energies, of emission of two photons
by an electron in the planar Doyle-Turner potential \cite{Baier1998,Doyle,avakian1982,Moller1995403}.
We do this by finding numerical solutions of the Dirac equation by
solving the problem in a basis of plane waves, which is possible due
to the periodicity of the transverse potential in a crystal, as shown
in \cite{Wistisen2019}. If the cascade terms are enough to properly
describe the radiation emission is a highly relevant question as it
closely relates to the phenomenon of quantum radiation reaction, the
emission of multiple photons when $\chi$ is large, \cite{PhysRevLett.105.220403},
recently studied using channeling radiation and in laser experiments
\cite{wistisen2018experimental,PhysRevX.8.031004,PhysRevX.8.011020}.
In the crystal experiment it was seen that even for energies as high
as $180$ GeV positrons, where it could be expected that the constant
field approximation would be acceptable, it was shown that discrepancies
arise due to this, and therefore a more general theory was called
for. The current theory of quantum radiation reaction in lasers relies
on the local constant field approximation \cite{PhysRevLett.105.220403,PhysRevLett.111.054802,PhysRevLett.112.015001,Baier1998,Vranic_2016,PhysRevLett.113.044801},
and it is unknown if one can calculate the emission of many photons
in a way that avoids calculating all the corresponding higher order
diagrams, when going beyond the constant field approximation. This
question will be addressed in the case of a crystal, in the current
paper.

We use the Feynman slash notation  such that $\slashed a=a_{\mu}\gamma^{\mu}$, where $\gamma^{\mu}$ are the Dirac  gamma matrices and $a^{\mu}$ an arbitrary four-vector. We adopt the metric tensor  $\eta^{\mu\nu}=\text{diag}(+1,-1,-1,-1)$.

\section{Formalism}

In QED the transition amplitude from a given initial state $\left|i\right\rangle $
to a final state $\left|f\right\rangle $ is given by

\begin{equation}
S_{fi}=\left\langle f\right|U(\infty,-\infty)\left|i\right\rangle \label{eq:Smat}
\end{equation}
where $U$ is the time evolution operator, often written as $U(\infty,-\infty)=\mathcal{T}\text{exp}\left(-i\int_{-\infty}^{\infty}V(t)dt\right)$
where $\mathcal{T}$ is the time-ordering operator and $V(t)=\int e\bar{\Psi}\slashed A\Psi d^{3}x$
is the quantized interaction. We then write our quantized fields as

\begin{equation}
\Psi=\sum_{s=1}^{2}\int\frac{d^{3}p}{(2\pi)^{3}}\left[b_{p}^{s}\psi_{p,s}^{-}(x)+c_{p}^{s\dagger}\psi_{p,s}^{+}(x)\right],\label{eq:spinoroperator}
\end{equation}

\begin{equation}
\slashed A=\int\frac{d^{3}k}{(2\pi)^{3}}\sqrt{\frac{4\pi}{2\omega}}\sum_{r=1}^{2}\left[\slashed\epsilon_{r}a_{k}^{r}e^{-ikx}+\slashed\epsilon_{r}^{*}a_{k}^{r\dagger}e^{ikx}\right],\label{eq:Aoperator}
\end{equation}
where $\psi_{p,s}^{-}(x)$ and $\psi_{p,s}^{+}(x)$ are an orthonormal
and complete set of electron and positron solutions, respectively,
in the background field. $\int\frac{d^{3}p}{(2\pi)^{3}}$ denotes
a summation over all states, and $p$ the relevant quantum numbers
which we will find later. The $b$, $c$ and $a$ operators are the
annihilation operators of the electron, positron and photon field
respectively, obeying the relations, that the only non-zero (anti-)commutators
are $\left\{ b_{p}^{r},b_{q}^{s\dagger}\right\} =\left\{ c_{p}^{r},c_{q}^{s\dagger}\right\} =[a_{p}^{r},a_{q}^{s\dagger}]=(2\pi)^{3}\delta^{rs}\delta^{(3)}(\boldsymbol{p}-\boldsymbol{q})$,
where the $\{\}$ brackets denote the anti-commutator and $[]$ the
commutator.

In \cite{Wistisen2019,PhysRevA.98.022131} we discussed the Dirac
equation with the potential found in the crystal, but we will here
repeat the results we need in order to calculate the emission of 2
photons. It was found in \cite{Wistisen2019} that the electron solution
can be written as follows

\begin{equation}
\psi^{-}(x)=\frac{1}{\sqrt{2\varepsilon}}e^{i(p_{x}x+p_{z}z-\varepsilon t)}U(y),\label{eq:electronwavefunc}
\end{equation}
and the positron solutions can then be written as (see appendix A)

\[
\psi^{+}(x)=\frac{1}{\sqrt{2\varepsilon}}e^{-i(p_{x}x+p_{z}z-\varepsilon t)}V(y)
\]
and $U$ and $V$ are given by

\[
U(y)=\sqrt{\varepsilon+m}\left(\begin{array}{c}
\boldsymbol{s}^{-}\\
\frac{\boldsymbol{\sigma}\cdot\tilde{\boldsymbol{p}}}{\varepsilon+m}\boldsymbol{s}^{-}
\end{array}\right)I^{-}(y)
\]

\[
V(y)=\sqrt{\varepsilon+m}\left(\begin{array}{c}
\frac{\boldsymbol{\sigma}\cdot\tilde{\boldsymbol{p}}}{\varepsilon+m}\boldsymbol{s}^{+}\\
\boldsymbol{s}^{+}
\end{array}\right)I^{+}(y)
\]
where $\tilde{\boldsymbol{p}}=\left(p_{x}+q\varphi(y),\text{sign}(q)i\frac{d}{dy},p_{z}\right)$,
$\varphi(y)$ is the electrostatic potential, $q=\pm e$ is the charge,
the superscript on $I(y)$ refers to the charge sign, $\boldsymbol{s}$
is a two component vector describing the spin, which we can choose
as either $\left(\begin{array}{cc}
1 & 0\end{array}\right)^{T}$ or $\left(\begin{array}{cc}
0 & 1\end{array}\right)^{T}$ , corresponding to spin-up and spin-down respectively for the electron,
and opposite for the positron. From the choice of the form of the
spinors $U$ and $V$, it is also clear that $\varepsilon$ positive
should be used (see appendix A). $I(y)$ is the solution
to the equation

\begin{equation}
\left[-\frac{1}{2\varepsilon}\frac{d^{2}}{dy^{2}}+q\varphi(y)\right]I(y)=\frac{\varepsilon^{2}-p_{x}^{2}-p_{z}^{2}-m^{2}}{2\varepsilon}I(y),\label{eq:schrodeq}
\end{equation}
For $\varphi(y)$ we will use the Doyle-Turner model \cite{Baier1998,Doyle,avakian1982,Moller1995403},
chosen as symmetric around 0. In a crystal this potential $\varphi(y)$
is periodic with the period of the inter planar distance which we
will denote as $d_{p}$. Because of this, the solution (for the electron)
can be written as a Bloch wave such that

\begin{equation}
I^{-}(y)=e^{ik_{B}^{-}y}u_{k_{B}}^{-}(y),\label{eq:Idef}
\end{equation}
and where $u_{k_{B}^{-}}^{-}(y)$ is also periodic with period $d_{p}$
and $k_{B}^{-}$ is the Bloch momentum, which can be taken to be in
the interval $0\leq k_{B}^{-}<k_{0}$, $k_{0}=\frac{2\pi}{d_{p}}$.
It then follows from Blochs theorem that these solutions form an orthogonal
and complete set of solutions of Eq. (\ref{eq:schrodeq}). Inserting
$I^{-}(y)$ of Eq. (\ref{eq:Idef}) into Eq. (\ref{eq:schrodeq})
gives us the equation governing $u_{k_{B}}^{-}(y)$

\begin{flalign}
\left(-\frac{1}{2\varepsilon}\left[\frac{d^{2}}{dy^{2}}+2ik_{B}\frac{d}{dy}-k_{B}^{2}\right]+q\varphi(y)\right)u_{k_{B}}^{-}(y)\nonumber \\
=\frac{\varepsilon^{2}-p_{x}^{2}-p_{z}^{2}-m^{2}}{2\varepsilon}u_{k_{B}}^{-}(y).\label{eq:ukbdiffeq}
\end{flalign}
The periodicity of $u_{k_{B}}(y)$ means it can be written as a Fourier
series,

\begin{equation}
u_{k_{B}}^{-}(y)=\sum_{j}c_{j}e^{ijk_{0}y},\label{eq:planewaveu}
\end{equation}
To ensure normalization we should have $\sum_{j}\left|c_{j}\right|^{2}=1$
(see appendix B). It is now clear that this is an eigenvalue
problem for each $k_{B}$ where the quantized eigenvalue is 
\begin{equation}
E_{n}=\frac{\varepsilon^{2}-p_{x}^{2}-p_{z}^{2}-m^{2}}{2\varepsilon},\label{eq:energyquant}
\end{equation}
where $n$ is the quantum number corresponding to the value of this
energy in ascending order and where $0$ is the ground state. This
equation leads to a quantization of e.g. $p_{x}$. The coefficients
$c_{j}$ are found by solving the matrix eigenvalue problem obtained
by inserting Eq. (\ref{eq:planewaveu}) in Eq. (\ref{eq:ukbdiffeq})
and multiply with $\frac{1}{d_{p}}e^{-ilk_{0}y}$ and integrate over
$y$ from $0$ to $d_{p}$ to exploit orthogonality

\begin{align}
 & \sum_{j}\frac{1}{2\varepsilon}\left[jk_{0}+k_{B}\right]^{2}\delta_{j,l}c_{j}\nonumber \\
 & +\sum_{j}c_{j}\frac{1}{d_{p}}\int q\varphi(y)e^{i(j-l)k_{0}y}dy\nonumber \\
 & =\sum_{j}\frac{\varepsilon^{2}-p_{x}^{2}-p_{z}^{2}-m^{2}}{2\varepsilon}\delta_{j,l}c_{j}.\label{eq:Matrixeq}
\end{align}
This was done with the electron function $I^{-}(y)$ in mind, but
the positron coefficients can be obtained just by changing $q$. With
these things taken into consideration, we now see that we can write
the $U(y)$ and $V(y)$ functions in terms of the coefficients $c_{j}$
such that

\begin{equation}
U(y)=\sum_{j}c_{j}^{-}\boldsymbol{S}_{j}^{-}e^{i(jk_{0}+k_{B})y},\label{eq:Uplanewave}
\end{equation}

\begin{equation}
V(y)=\sum_{j}c_{j}^{+}\boldsymbol{S}_{j}^{+}e^{-i(jk_{0}+k_{B})y},\label{eq:Uplanewave-1}
\end{equation}
where

\begin{equation}
\boldsymbol{S}_{j}^{-}=\sqrt{\varepsilon+m}\left(\begin{array}{c}
\boldsymbol{s}^{-}\\
\frac{\boldsymbol{\sigma}\cdot\boldsymbol{p}_{j}}{\varepsilon+m}\boldsymbol{s}^{-}
\end{array}\right),\label{eq:bigS}
\end{equation}

\begin{equation}
\boldsymbol{S}_{j}^{+}=\sqrt{\varepsilon+m}\left(\begin{array}{c}
\frac{\boldsymbol{\sigma}\cdot\boldsymbol{p}_{j}}{\varepsilon+m}\boldsymbol{s}^{+}\\
\boldsymbol{s}^{+}
\end{array}\right),\label{eq:bigS-1}
\end{equation}
where $\boldsymbol{p}_{j}=(p_{x}+E_{n}-\frac{(jk_{0}+k_{B})^{2}}{2\varepsilon},jk_{0}+k_{B},p_{z})$.
For the calculation of radiation emission from electrons we will need
the quantity $\bar{\boldsymbol{S}}_{f}^{-}\slashed{\epsilon}^{*}\boldsymbol{S}_{i}^{-}$,
where we have put labels for the initial state $i$ and final state
$f$, however these still each depend on the index $j$. This quantity
can then be written as

\begin{equation}
\bar{\boldsymbol{S}}_{f}^{-}\slashed{\epsilon}^{*}\boldsymbol{S}_{i}^{-}=-\boldsymbol{s}_{f}^{T}\left[\boldsymbol{\epsilon}\cdot\boldsymbol{A}+i\boldsymbol{B}\cdot\boldsymbol{\sigma}\right]\boldsymbol{s}_{i}\label{eq:Srelation}
\end{equation}
where 

\begin{equation}
\boldsymbol{A}=\sqrt{\frac{\varepsilon_{f}+m}{\varepsilon_{i}+m}}\boldsymbol{p}_{i}+\sqrt{\frac{\varepsilon_{i}+m}{\varepsilon_{f}+m}}\boldsymbol{p}_{f},\label{eq:bigA}
\end{equation}

\begin{equation}
\boldsymbol{B}=\boldsymbol{\epsilon}^{*}\times\left(\sqrt{\frac{\varepsilon_{f}+m}{\varepsilon_{i}+m}}\boldsymbol{p}_{i}-\sqrt{\frac{\varepsilon_{i}+m}{\varepsilon_{f}+m}}\boldsymbol{p}_{f}\right).\label{eq:bigB}
\end{equation}
Now since we have an orthonormal complete set of solutions, we can
write the propagator in terms of these states as \cite{beresteckij_quantum_2008}

\begin{flalign*}
G(x_{2},x_{1}) & =\int\frac{d^{3}p}{(2\pi)^{3}}\sum_{n,s}\\
 & \theta(t_{2}-t_{1})e^{-i\varepsilon(t_{2}-t_{1})}\psi_{p,n,s}^{-}(\boldsymbol{x}_{2})\bar{\psi}_{p,n,s}^{-}(\boldsymbol{x}_{1})\\
 & -\theta(t_{1}-t_{2})e^{i\varepsilon(t_{2}-t_{1})}\psi_{p,n,s}^{+}(\boldsymbol{x}_{2})\bar{\psi}_{p,n,s}^{+}(\boldsymbol{x}_{1}).
\end{flalign*}
This expression can be simplified due to the simple expression for
the wave functions in all coordinates but the $y$ coordinate. However,
we will not carry this out, as it is easier to see how the cascade
part of the radiation emission arises by starting from the above expression.

\section{Single photon emission and cascade}

We will now briefly mention some results obtained in \cite{Wistisen2019}
on the single photon emission probability which is relevant to build
the expected cascade contribution. We found that the rate of emission
is given by

\begin{flalign}
dW_{i\rightarrow f}^{(1)} & =\frac{1}{(2\pi)^{2}}\left|\mathcal{M}_{i\rightarrow f}\right|^{2}\delta(\varepsilon_{f}+\omega-\varepsilon_{i})d^{3}k,\label{eq:Onephotemission}
\end{flalign}
where we defined

\begin{equation}
\mathcal{M}_{i\rightarrow f}=e\sqrt{\frac{4\pi}{2\omega}}\frac{1}{2\sqrt{\varepsilon_{f}\varepsilon_{i}}}\sum_{j}c_{n_{B}+j,f}^{*}c_{j,i}\bar{\boldsymbol{S}}_{n_{B}+j,f}^{-}\slashed{\epsilon}^{*}\boldsymbol{S}_{j,i}^{-},\label{eq:Mtilde}
\end{equation}
where $n_{B}$ is the integer such that $0\leq k_{B,f}<k_{0}$, where
$k_{B,f}=k_{B,i}-k_{y}-n_{B}k_{0}$, $\boldsymbol{S}_{j,i}$ corresponds
to the initial state and $c_{j,i}$ is coefficient with index $j$
corresponding to the initial state $i$. See the appendix
of \cite{Wistisen2019} for the details on why $\mathcal{M}$ reduces
to a single sum over $j$. As shown in \cite{Wistisen2019} there
are large terms in $\varepsilon_{f}+\omega-\varepsilon_{i}$ which
cancel, leaving behind the relevant small terms, because the relevant
transverse energies $E_{n}$, comparable to the potential depth, are
much smaller than the whole particle energy i.e. eV versus GeV. We
could rewrite the content of the delta function as

\begin{flalign}
f(\theta) & =\varepsilon_{f}+\omega-\varepsilon_{i}\nonumber \\
 & \simeq E_{n_{f}}-E_{n_{i}}+\frac{m^{2}}{2\varepsilon_{f}}-\frac{m^{2}}{2\varepsilon_{i}}+\frac{\omega\theta^{2}}{2}\left(1+\frac{\omega\text{sin}^{2}\varphi}{\varepsilon_{f}}\right).\label{eq:ftheta}
\end{flalign}
Now we may use that $\delta(\varepsilon_{f}+\omega-\varepsilon_{i})=\frac{1}{|f'(\theta_{0})|}\delta(\theta-\theta_{0})$
where $\theta_{0}$ is the positive solution to $f(\theta)=0$. From
the formula for single photon emission, Eq. (\ref{eq:Onephotemission}),
we can construct the cascade contribution to two photon emission.
We wish to know the probability of finding a photon in the momentum
interval $d^{3}k_{1}$ around $k_{1}$ while also finding a photon
within another interval $d^{3}k_{2}$ around $k_{2}$. This can happen
in two ways, either the particle emits $k_{1}$ while transitioning
from the initial state, and then subsequently $k_{2}$ or vice versa.
We are however interested in the angular integrated spectrum, that
is $dP_{i\rightarrow f}^{(\text{cascade})}/d\omega_{1}d\omega_{2}$
and therefore an additional factor of $\frac{1}{2}$ must be added
due to counting the same point in phase space twice \cite{feynman1965feynman},
and so we obtain

\begin{flalign}
\frac{dP_{i\rightarrow f}^{(\text{cascade})}}{d\omega_{1}d\omega_{2}} & =\frac{T^{2}}{2}\sum_{v}\frac{1}{2}\left[\frac{dW_{i\rightarrow v}^{(1)}}{d\omega}(\omega_{1})\frac{dW_{v\rightarrow f}^{(1)}}{d\omega}(\omega_{2})\right.\nonumber \\
 & \left.+\frac{dW_{i\rightarrow v}^{(1)}}{d\omega}(\omega_{2})\frac{dW_{v\rightarrow f}^{(1)}}{d\omega}(\omega_{1})\right].\label{eq:cascadeconstruct}
\end{flalign}

\section{Two photon emission}

Expanding the time evolution operator to second order, allowing for
two photon emission we have that the S-matrix element is

\begin{flalign}
S_{i\rightarrow f}^{(2)} & =-\left\langle f\right|\frac{1}{2}\iintop_{-\infty}^{\infty}\mathcal{T}\hat{V}(t_{2})\hat{V}(t_{1})dt_{1}dt_{2}\left|i\right\rangle .
\end{flalign}
When specifying the final state as $\left\langle p_{f},k_{1},k_{2}\right|$,
an electron and two photons and the initial state as just an electron,
$\left|p_{i},0,0\right\rangle $, $S_{fi}^{(2)}$ can be rewritten
in terms of the wave functions and the propagator. In \cite{beresteckij_quantum_2008}
this is done for the Compton scattering matrix element, which is the
same diagram as here, except that an incoming photon is instead outgoing.
The matrix element is therefore

\begin{flalign}
S_{i\rightarrow f}^{(2)} & =-ie^{2}\sqrt{\frac{4\pi}{2\omega_{1}}}\sqrt{\frac{4\pi}{2\omega_{2}}}\iint d^{4}x_{2}d^{4}x_{1}\nonumber \\
 & \times\bar{\psi}_{f}^{-}(x_{2})\slashed{\epsilon}_{2}^{*}e^{ik_{2}x_{2}}G(x_{2},x_{1})\slashed{\epsilon}_{1}^{*}e^{ik_{1}x_{1}}\psi_{i}^{-}(x_{1})\nonumber \\
 & +(\epsilon_{1},k_{1})\leftrightarrow(\epsilon_{2},k_{2}).\label{eq:Sf2}
\end{flalign}
Now we define

\begin{align}
 & M_{i\rightarrow v}^{-}=e\sqrt{\frac{4\pi}{2\omega_{1}}}\int d^{3}\boldsymbol{x}\bar{\psi}_{v}^{-}(\boldsymbol{x})\slashed{\epsilon}_{1}^{*}e^{-i\boldsymbol{k}_{1}\cdot\boldsymbol{x}}\psi_{i}^{-}(\boldsymbol{x})\nonumber \\
 & =(2\pi)^{3}\delta(p_{x,i}-p_{x,v}-k_{x})\delta(p_{z,i}-k_{z,1}-p_{z,v})\nonumber \\
 & \times\delta(k_{B,i}-k_{y,1}-k_{B,v}-n_{B,1}k_{0})e\sqrt{\frac{4\pi}{2\omega_{1}}}\frac{1}{2\sqrt{\varepsilon_{i}\varepsilon_{v}}}\nonumber \\
 & \times\sum_{j}c_{n_{B}+j,v}^{*}c_{j,i}\bar{\boldsymbol{S}}_{n_{B}+j,v}^{-}\slashed{\epsilon}_{1}^{*}\boldsymbol{S}_{j,i}^{-}\nonumber \\
 & =(2\pi)^{3}\delta(p_{x,i}-p_{x,v}-k_{x})\delta(p_{z,i}-k_{z,1}-p_{z,v})\nonumber \\
 & \times\delta(k_{B,i}-k_{y,1}-k_{B,v}-n_{B,1}k_{0})\times\mathcal{M}_{i\rightarrow v}^{-}(\boldsymbol{k}_{1},\epsilon_{1}),\label{eq:M++}
\end{align}
where $\mathcal{M}$ is defined as in Eq. (\ref{eq:Mtilde}) where
$v$ is used to denote the virtual state from the propagator, and
is shorthand for the dependence on $p_{x,v}$, $k_{B,v}$, $p_{z,v}$,
$n_{v}$ and $s_{v}$. The superscript $-$ on $M^{-}$ and $\mathcal{M}^{-}$
denotes that the virtual state is the electron state $\psi_{v}^{-}$,
and $M^{+}$, $\mathcal{M}^{+}$ is the same but with the positron
virtual state. The matrix element may then be written as

\begin{flalign}
S_{fi}^{(2)} & =i\iint dt_{2}dt_{1}\sum_{n_{v},s_{v}}\int\frac{d^{3}p_{v}}{(2\pi)^{3}}\nonumber \\
 & -\theta(t_{2}-t_{1})e^{i\left(\omega_{1}+\varepsilon_{v}-\varepsilon_{i}\right)t_{1}}e^{i\left(\omega_{2}+\varepsilon_{f}-\varepsilon_{v}\right)t_{2}}M_{i\rightarrow v}^{-}M_{v\rightarrow f}^{-}\nonumber \\
 & \theta(t_{1}-t_{2})e^{i\left(\omega_{1}-\varepsilon_{i}-\varepsilon_{v}\right)t_{1}}e^{i\left(\omega_{2}+\varepsilon_{f}+\varepsilon_{v}\right)t_{2}}M_{i\rightarrow v}^{+}M_{v\rightarrow f}^{+}.\label{eq:SintermsofM}
\end{flalign}
Therefore the term in the second line is seen as the electron first
emits a photon with momentum $\boldsymbol{k}_{1}$ at $t_{1}$ and
then propagates to a later time $t_{2}$ and emits a second photon
with momentum $\boldsymbol{k}_{2}$. The term in the third line is
then the electron emitting the photon with momentum $\boldsymbol{k}_{1}$
at a time $t_{1}$ turning the electron into a positron going into
the past and emitting the photon with momentum $\boldsymbol{k}_{2}$
at the earlier time $t_{2}$. This last term is heavily suppressed
in our case which we can see as follows. Denote $a=\varepsilon_{v}+\omega_{1}-\varepsilon_{i}$
and $b=\varepsilon_{f}+\omega_{2}-\varepsilon_{v}$, then we may use
that

\begin{equation}
\theta(t_{1}-t_{2})=\frac{i}{2\pi}\int_{-\infty}^{\infty}\frac{1}{\varepsilon_{V}+i\epsilon}e^{i(t_{2}-t_{1})\varepsilon_{V}}d\varepsilon_{V}
\end{equation}
where $\epsilon$ is a small real number for which one in the end
should take the limit $\epsilon\rightarrow0$ and therefore we have
\begin{flalign}
 & \iint_{-\infty}^{\infty}dt_{2}dt_{1}\theta(t_{1}-t_{2})e^{iat_{2}}e^{ibt_{1}}\nonumber \\
 & =2\pi i\delta(a+b)\frac{1}{-a+i\epsilon}\nonumber \\
 & =2\pi i\delta(\varepsilon_{f}+\omega_{1}+\omega_{2}-\varepsilon_{i})\frac{1}{\varepsilon_{i}-\varepsilon_{v}-\omega_{1}+i\epsilon}.\label{eq:Propagator time}
\end{flalign}
We have also that $-\theta(t_{2}-t_{1})=\frac{i}{2\pi}\int\frac{1}{\varepsilon_{V}-i\epsilon}e^{i(t_{2}-t_{1})\varepsilon_{V}}d\varepsilon_{V}$
and therefore we have the term from the third line of Eq. (\ref{eq:SintermsofM})
carries the factor of

\begin{equation}
2\pi i\delta(\varepsilon_{f}+\omega_{1}+\omega_{2}-\varepsilon_{i})\frac{1}{\varepsilon_{i}+\varepsilon_{v}-\omega_{1}-i\epsilon},
\end{equation}
and therefore this term will always be very far off-shell, as the
virtual particle on-shell condition can never be met as it corresponds
to the spontaneous production of an electron, positron and photon
from the crystal field, where the produced positron is subsequently
annihilated with the incoming electron to emit another photon. Having
carried out the integrations over time we obtain that

\begin{flalign}
S_{fi}^{(2)} & =-\sum_{n_{v},s_{v}}\int\frac{d^{3}p_{v}}{(2\pi)^{3}}2\pi\delta(\varepsilon_{f}+\omega_{1}+\omega_{2}-\varepsilon_{i})\nonumber \\
 & \left(\frac{M_{i\rightarrow v}^{-}M_{v\rightarrow f}^{-}}{\varepsilon_{i}-\varepsilon_{v_{1}}-\omega_{1}-i\epsilon}\right.\nonumber \\
 & +\frac{M_{i\rightarrow v}^{+}M_{v\rightarrow f}^{+}}{\varepsilon_{i}+\varepsilon_{v_{1}}-\omega_{1}+i\epsilon}\nonumber \\
 & \left.+(\epsilon_{1},k_{1})\leftrightarrow(\epsilon_{2},k_{2})\right).
\end{flalign}
Now we may integrate over $p_{v}$ to obtain
\begin{flalign*}
S_{fi}^{(2)} & =-\sum_{n_{v},s_{v}}\\
 & \left(\frac{\mathcal{M}_{i\rightarrow v_{1}}^{-}\mathcal{M}_{v_{1}\rightarrow f}^{-}}{\varepsilon_{i}-\varepsilon_{v_{1}}-\omega_{1}-i\epsilon}\right.\\
 & \left.+\frac{\mathcal{M}_{i\rightarrow v_{1}}^{+}\mathcal{M}_{v_{1}\rightarrow f}^{+}}{\varepsilon_{i}+\varepsilon_{v_{1}}-\omega_{1}+i\epsilon}\right)\\
 & \times(2\pi)^{4}\delta(\varepsilon_{f}+\omega_{1}+\omega_{2}-\varepsilon_{i})\delta(p_{x,i}-k_{x,1}-k_{x,2}-p_{x,f})\\
 & \times\delta(p_{z,i}-k_{z,1}-k_{z,2}-p_{z,f})\\
 & \times\delta(k_{B,i}-k_{y,1}-k_{y,2}-k_{B,f}-\left(n_{B,1}+n_{B,2}\right)k_{0})\\
 & +(\epsilon_{1},k_{1})\leftrightarrow(\epsilon_{2},k_{2}),
\end{flalign*}
and then $v_{1}$ denotes the virtual state with momentum given by
$p_{x,v_{1}}=p_{x,i}-k_{x,1}$, $p_{z,v_{1}}=p_{z,i}-k_{z,1}$ and
$k_{B,v_{1}}^{-}=k_{B,i}^{-}-k_{y,1}-n_{B,1}^{-}k_{0}$ and $-k_{B,v_{1}}^{+}=k_{B,i}^{-}-k_{y,1}-n_{B,1}^{+}k_{0}$,
i.e. that photon with label $1$ is emitted at the vertex connected
with the initial particle. From the amplitude we get the transition
probability according to

\begin{flalign}
dP^{(2)} & =\frac{1}{2}\int\sum_{n_{f},s_{f}}|S_{fi}^{(2)}|^{2}\frac{dp_{x,f}dk_{B,f}dp_{z,f}}{(2\pi)^{3}}\frac{d^{3}k_{1}}{(2\pi)^{3}}\frac{d^{3}k_{2}}{(2\pi)^{3}},\nonumber \\
 & =\sum_{n_{f},s_{f}}\left|\sum_{n_{v},s_{v}}\frac{\mathcal{M}_{i\rightarrow v_{1}}^{-}\mathcal{M}_{v_{1}\rightarrow f}^{-}}{\varepsilon_{i}-\varepsilon_{v_{1}}-\omega_{1}-i\epsilon}\right.\nonumber \\
 & +\frac{\mathcal{M}_{i\rightarrow v_{1}}^{+}\mathcal{M}_{v_{1}\rightarrow f}^{+}}{\varepsilon_{i}+\varepsilon_{v_{1}}-\omega_{1}+i\epsilon}\nonumber \\
 & \left.+(\epsilon_{1},k_{1})\leftrightarrow(\epsilon_{2},k_{2})\right|^{2}\nonumber \\
 & \times\frac{T}{(2\pi)^{5}}\delta(\varepsilon_{f}+\omega_{1}+\omega_{2}-\varepsilon_{i})d^{3}k_{1}d^{3}k_{2},\label{eq:Fullprob}
\end{flalign}
where we have added a factor of $1/2$ in front due to identical particles
in the final state, and that we in the end want to integrate over
all angles, and would therefore, again, be counting double \cite{feynman1965feynman}.
From this full result, it is seen that the result can diverge when
$\epsilon\rightarrow0$ because $\varepsilon_{i}-\varepsilon_{v}-\omega_{1}=0$
is possible. The nature of the divergence is however different for
some of the terms, namely the ones which are the norm square of each
term underneath the sum, $\left|\mathcal{M}_{i\rightarrow v_{1}}^{-}\mathcal{M}_{v_{1}\rightarrow f}^{-}/\left(\varepsilon_{i}-\varepsilon_{v_{1}}-\omega_{1}-i\epsilon\right)\right|^{2}$,
where the limit of $\epsilon\rightarrow0$ will yield an infinite
result, even after integration over one of the angles $\theta_{1}$
or $\theta_{2}$. On the other hand, while the remaining terms, of
the interference type, still diverge, they can be integrated over
$\theta_{1}$ or $\theta_{2}$ to yield a convergent result. To learn
the meaning of this divergence due to the denominator, see also \cite{hu2011multi},
we may write

\begin{equation}
\left|\frac{1}{b-i\epsilon}\right|^{2}=\frac{1}{b^{2}+\epsilon^{2}},
\end{equation}
and note that 
\begin{equation}
\underset{\epsilon\rightarrow0}{\text{lim}}\frac{\epsilon}{b^{2}+\epsilon^{2}}=\pi\delta(b).\label{eq:deltafunc}
\end{equation}
if we evaluate the integrals of $\mathcal{M}_{i\rightarrow v_{1}}^{-}\mathcal{M}_{v_{1}\rightarrow f}^{-}$
with the factor $\delta(a+b)\delta(b)$ we get well defined results,
as this just amounts to the product of two 1.st order emissions. It
is therefore useful to write

\begin{equation}
\left|\frac{1}{b-i\epsilon}\right|^{2}=\frac{1}{\epsilon}\frac{\epsilon}{b^{2}+\epsilon^{2}},\label{eq:deltarewrite}
\end{equation}
where then the factor $\epsilon/(b^{2}+\epsilon^{2})$ acts like a
delta-function for small enough $\epsilon$, yielding a finite value
when we perform the integrals in Eq. (\ref{eq:Fullprob}), and then
it is clear that this is divergent as $\epsilon\rightarrow0$ due
to the factor of $1/\epsilon$. However this should be understood
in terms of an additional factor of $T$ for this term. To see this,
consider the origin of this expression from Eq. (\ref{eq:Propagator time}),
but consider instead that we had a finite time, and integrate over
$a$ and $b$

\begin{flalign}
 & \int\left|\iint_{0}^{T}\theta(t_{1}-t_{2})e^{iat_{2}}e^{ibt_{1}}dt_{1}dt_{2}\right|^{2}dadb\nonumber \\
 & =(2\pi)^{2}\int_{0}^{T}\theta(t_{1}-t_{2})dt_{1}dt_{2}=(2\pi)^{2}\frac{T^{2}}{2},\label{eq:integralofG}
\end{flalign}
and we also have that
\begin{flalign}
 & \int\left|2\pi i\delta(a+b)\frac{1}{b-i\epsilon}\right|^{2}dadb\nonumber \\
 & =2\pi T\frac{\pi}{\epsilon},\label{eq:integralofGeps}
\end{flalign}
and so we see that we must replace $\left|\frac{1}{b-i\epsilon}\right|^{2}\rightarrow\pi T\delta(b)$,
and therefore these terms turn out to give us the cascade contribution.
To see how the probability from Eq. (\ref{eq:Fullprob}) splits up
into this cascade along with additional terms, we will denote the
quantity underneath the norm-square as $R^{-}=\sum_{n_{v}}J_{1}^{-}(n_{v})+J_{2}^{-}(n_{v})$
corresponding to the terms with the virtual electron and similarly
$R^{+}=\sum_{n_{v}}J_{1}^{+}(n_{v})+J_{2}^{+}(n_{v})$, where

\begin{equation}
J_{1}^{-}(n_{v})=\sum_{s_{v}}\frac{\mathcal{M}_{i\rightarrow v_{1}}^{-}\mathcal{M}_{v_{1}\rightarrow f}^{-}}{\varepsilon_{i}-\varepsilon_{v_{1}}-\omega_{1}-i\epsilon}\label{eq:Jplus}
\end{equation}

\begin{equation}
J_{1}^{+}(n_{v})=\sum_{s_{v}}\frac{\mathcal{M}_{i\rightarrow v_{1}}^{+}\mathcal{M}_{v_{1}\rightarrow f}^{+}}{\varepsilon_{i}+\varepsilon_{v_{1}}-\omega_{1}+i\epsilon},\label{eq:Jminus}
\end{equation}
and $J_{2}$ is $J_{1}$ with $(\epsilon_{1},k_{1})\leftrightarrow(\epsilon_{2},k_{2})$
and we then define $R=R^{+}+R^{-}$. The quantity we want is then
$|R|^{2}=|R^{+}|^{2}+|R^{-}|^{2}+2\text{Re}\left[R^{+}\left(R^{-}\right)^{*}\right]$.
In the $R^{+}$ term, it is never possible for the denominator to
become $0$ and therefore it can be directly calculated (see appendix C). For 
\begin{flalign}
|R^{-}|^{2} & =\left(\sum_{n_{v}}J_{1}^{-}(n_{v})+J_{2}^{-}(n_{v})\right)\nonumber \\
 & \times\left(\sum_{n_{v}'}J_{1}^{-}(n_{v}')+J_{2}^{-}(n_{v}')\right)^{*},
\end{flalign}
the product of the terms with the same subscript and where $n_{v}=n'_{v}$
are the cascade which are the only problematic terms and so need special
attention as described above. Therefore it is useful to employ that
\begin{flalign}
|R^{-}|^{2} & =\sum_{n_{v}}\left[J_{1}^{-}(n_{v})\left(R^{-}-J_{1}^{-}(n_{v})\right)^{*}\right.\nonumber \\
 & +J_{2}^{-}(n_{v})\left(R^{-}-J_{2}^{-}(n_{v})\right)^{*}\nonumber \\
 & \left.+|J_{1}^{-}(n_{v})|^{2}+|J_{2}^{-}(n_{v})|^{2}\right],
\end{flalign}
and so the terms in the first two lines are convergent contributions
to the one-step process and the terms on the last line are the cascade
terms, except that the spin sum is still underneath the norm-square.
In appendix D we show that the interference due to
spin will be $0$ when the photon polarization can be taken as real
and that either the sum over initial or final spins (we will do both)
is carried out. And so we can write the differential probability of
emission, with a given initial state, as

\begin{flalign}
dP^{(2)} & =\frac{1}{2}\frac{T}{(2\pi)^{5}}\delta(\varepsilon_{f}+\omega_{1}+\omega_{2}-\varepsilon_{i})d^{3}k_{1}d^{3}k_{2}\sum_{n_{f}}\nonumber \\
 & |R^{+}|^{2}+2\text{Re}\left[R^{+}\left(R^{-}\right)^{*}\right]\nonumber \\
 & +\sum_{n_{v}}\left\{ \sum_{s_{v}}\left[|\mathcal{M}_{i\rightarrow v_{1}}^{-}\mathcal{M}_{v_{1}\rightarrow f}^{-}|^{2}\pi T\delta(b)\right]\right.\nonumber \\
 & +J_{1}^{-}(n_{v})\left(R^{-}-J_{1}^{-}(n_{v})\right)^{*}\nonumber \\
 & \left.+(\epsilon_{1},k_{1})\leftrightarrow(\epsilon_{2},k_{2})\right\} .\label{eq:Fullprob-1}
\end{flalign}

\section{Choice of regularization}

Consider the terms proportional to $T^{2}$ from the above equation
\begin{figure}[t]
\includegraphics[width=1\columnwidth]{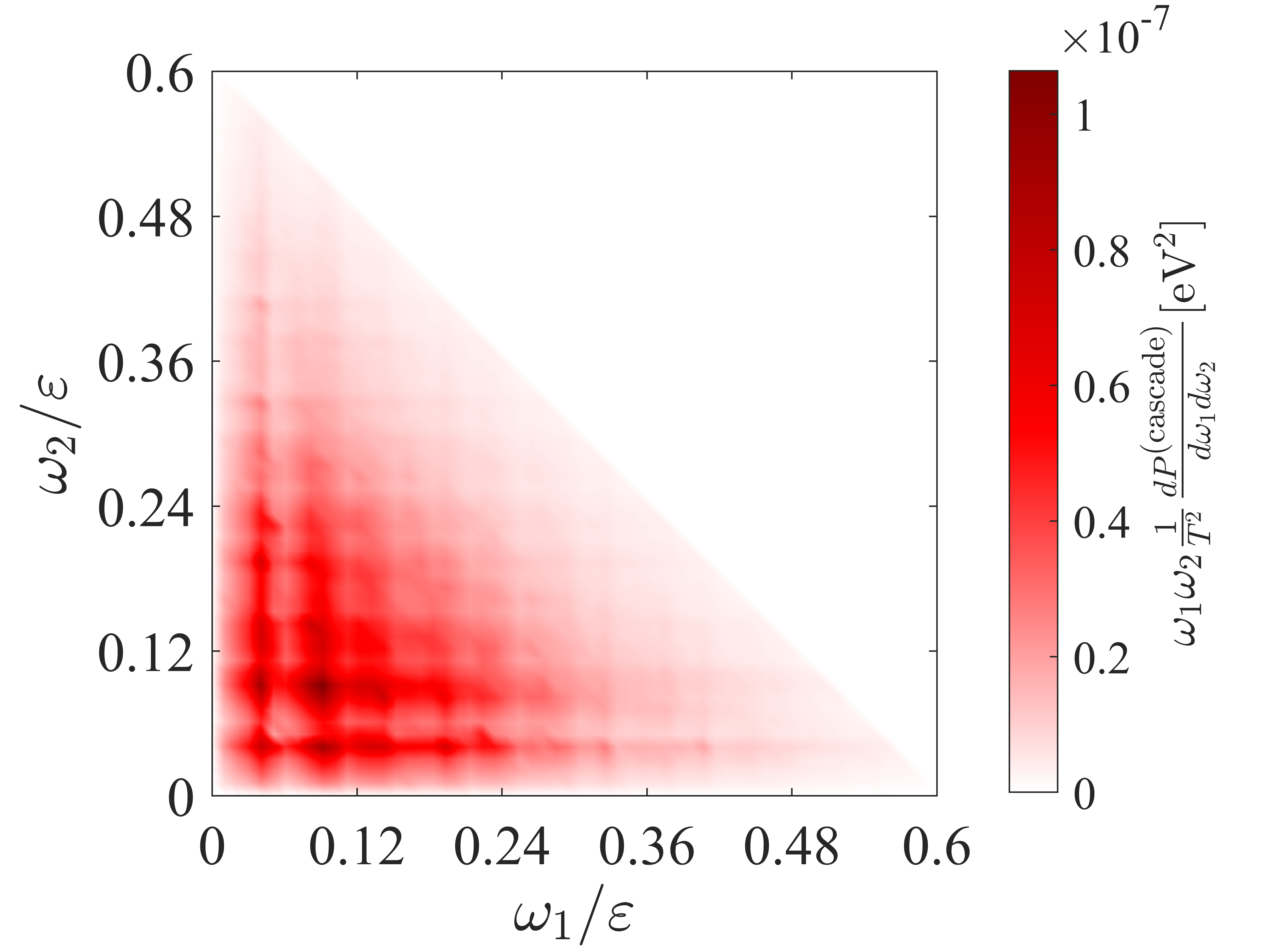}

\caption{The differential emission probability of two photons with energy $\omega_{1}$
and $\omega_{2}$ for the cascade contribution divided by $T^{2}$,
for the case mentioned in the text.\label{fig:Cascadeinf}}
\end{figure}
\begin{align}
dP^{\text{(cascade)}} & =\frac{1}{2}\frac{1}{(2\pi)^{4}}\frac{T^{2}}{2}d^{3}k_{1}d^{3}k_{2}\sum_{n_{f}}\nonumber \\
 & \sum_{n_{v},s_{v}}|\mathcal{M}_{i\rightarrow v_{1}}^{-}\mathcal{M}_{v_{1}\rightarrow f}^{-}|^{2}\delta(a+b)\delta(b)\nonumber \\
 & +(\epsilon_{1},k_{1})\leftrightarrow(\epsilon_{2},k_{2}),\label{eq:cascade}
\end{align}
which by comparison with Eq. (\ref{eq:cascadeconstruct}) and Eq.
(\ref{eq:Onephotemission}) is seen to be in agreement with the expected
cascade result. Above we chose a certain way to regularize the divergence,
by recognizing that the divergent terms correspond to the cascade
terms, and that in taking the time limit from $\pm\infty$, some information
about the duration of interaction was lost, which we put back in,
in a way that is correct when $T$ is large enough i.e. larger than
the photon formation length roughly estimated by $l_{f}=2\gamma^{2}(1-\omega/\varepsilon)/\omega$
\cite{Baier1998}, which in our case is roughly $\gamma/m\sim0.8$
\textmu m, because $\omega$ is on the order of $\varepsilon$. Another
way often found in literature \cite{oleinik1967,oleinik1968,roshchupkin1996resonant,PhysRevLett.98.043002,PhysRevD.90.043014}
is to say that the virtual state is unstable and therefore replace
the energy of the virtual particle according to $\varepsilon_{v}\rightarrow\varepsilon_{v}-i\Gamma_{v}/2$
where $\Gamma_{v}=\sum_{f}W_{v\rightarrow f}$ is the total decay
width of the virtual state from all processes. This is equivalent
to adding the effect of the line width in atomic Raman scattering
\cite{bransden2003physics}. Effectively this corresponds to replacing
the $\epsilon$ in the denominator with $-\Gamma_{v}/2$ which lifts
the divergence. However one can see that with this substitution, see
Eq. (\ref{eq:deltarewrite}), one would obtain that

\begin{equation}
\left|\frac{1}{b+i\Gamma_{v}/2}\right|^{2}=\frac{2\pi}{\Gamma_{v}}f(b),
\end{equation}
where $f(b)$ is a function peaked around $b=0$ which obeys $\int f(b)db=1$
and therefore resembles the delta-function $\delta(b)$, but with
a non-zero width $\Gamma_{v}$. If we then again calculate the cascade
part according to this we would obtain

\begin{align}
dP^{\text{(cascade)}*} & =\frac{1}{2}\frac{1}{(2\pi)^{5}}Td^{3}k_{1}d^{3}k_{2}\sum_{n_{f}}\nonumber \\
 & \sum_{n_{v},s_{v}}|\mathcal{M}_{i\rightarrow v_{1}}^{-}\mathcal{M}_{v_{1}\rightarrow f}^{-}|^{2}\delta(a+b)\frac{2\pi}{\Gamma_{v_{1}}}f(b)\nonumber \\
 & +(\epsilon_{1},k_{1})\leftrightarrow(\epsilon_{2},k_{2}),\label{eq:cascade-1}
\end{align}
and if we assume that the dominant contribution to the decay width
is due to radiation emission we have that the total width is
\[
\Gamma_{v}=\sum_{f}\int\frac{1}{(2\pi)^{2}}\left|\mathcal{M}_{v\rightarrow f}\right|^{2}\delta(\varepsilon_{f}+\omega-\varepsilon_{v})d^{3}k,
\]
therefore if we approximate $f(b)\simeq\delta(b)$ and integrate over
$d^{3}k_{2}$ and sum over $n_{f}$ we will obtain a factor of the
total rate $\Gamma_{v_{1}}$, which cancels out, and so we have that

\begin{align}
dP^{\text{(cascade)}*} & =\frac{1}{2}\frac{1}{(2\pi)^{2}}Td^{3}k_{1}\nonumber \\
 & \sum_{n_{v},s_{v}}|\mathcal{M}_{i\rightarrow v_{1}}^{-}|^{2}\delta(a)\nonumber \\
 & +(\epsilon_{1},k_{1})\leftrightarrow(\epsilon_{2},k_{2}),\label{eq:cascade-1-1}
\end{align}
which is just the single photon emission probability. Therefore this
approach leads to the prediction that it is just as likely to emit
2 photons as it is 1. This is not a meaningful result and the reason
is that the integration over time has been carried out over all times,
i.e. it is assumed that $T\gg1/\Gamma_{v}$ which means it is guaranteed
that the virtual state decays. However in that case not only 2 photon
emission is likely, also larger number of photons, which we do not
take into account. For Raman scattering the approach is reasonable
when $T\gg1/\Gamma_{v}$ such that it is guaranteed that an excited
state will decay before the observation is made. However if the interaction
time is very short $T\ll1/\Gamma_{v}$, it is also expected that Raman
scattering should have a dependence as $T^{2}$, as each sub process,
excitation and decay, is characterized by a rate, and the probability
is therefore the product of $\left(W^{\text{excite}}T\right)\left(W^{\text{decay}}T\right)$.
The substitution $\varepsilon_{v}\rightarrow\varepsilon_{v}-i\Gamma_{v}/2$
therefore corresponds to the replacement $W^{\text{decay}}T\rightarrow1$
and then combines the processes corresponding to the first order diagrams
of excitation first, and subsequently decay, with the second order
diagram which allows for off-resonant excitation and decay. We are
interested in the case when $T<1/\Gamma_{v}$ such that 2 photon emission
is unlikely compared to 1 photon emission, and therefore higher number
of photon emissions can be neglected. In this case one can also think
of the previously obtained result for the cascade contribution, as
the contribution of the finite crystal length to the line width, which
corresponds to setting $\Gamma_{v}/2=1/T$, which will be the dominant
contribution to the line width when $T\ll1/\Gamma_{v}$.
\begin{figure}[t]
\includegraphics[width=1\columnwidth]{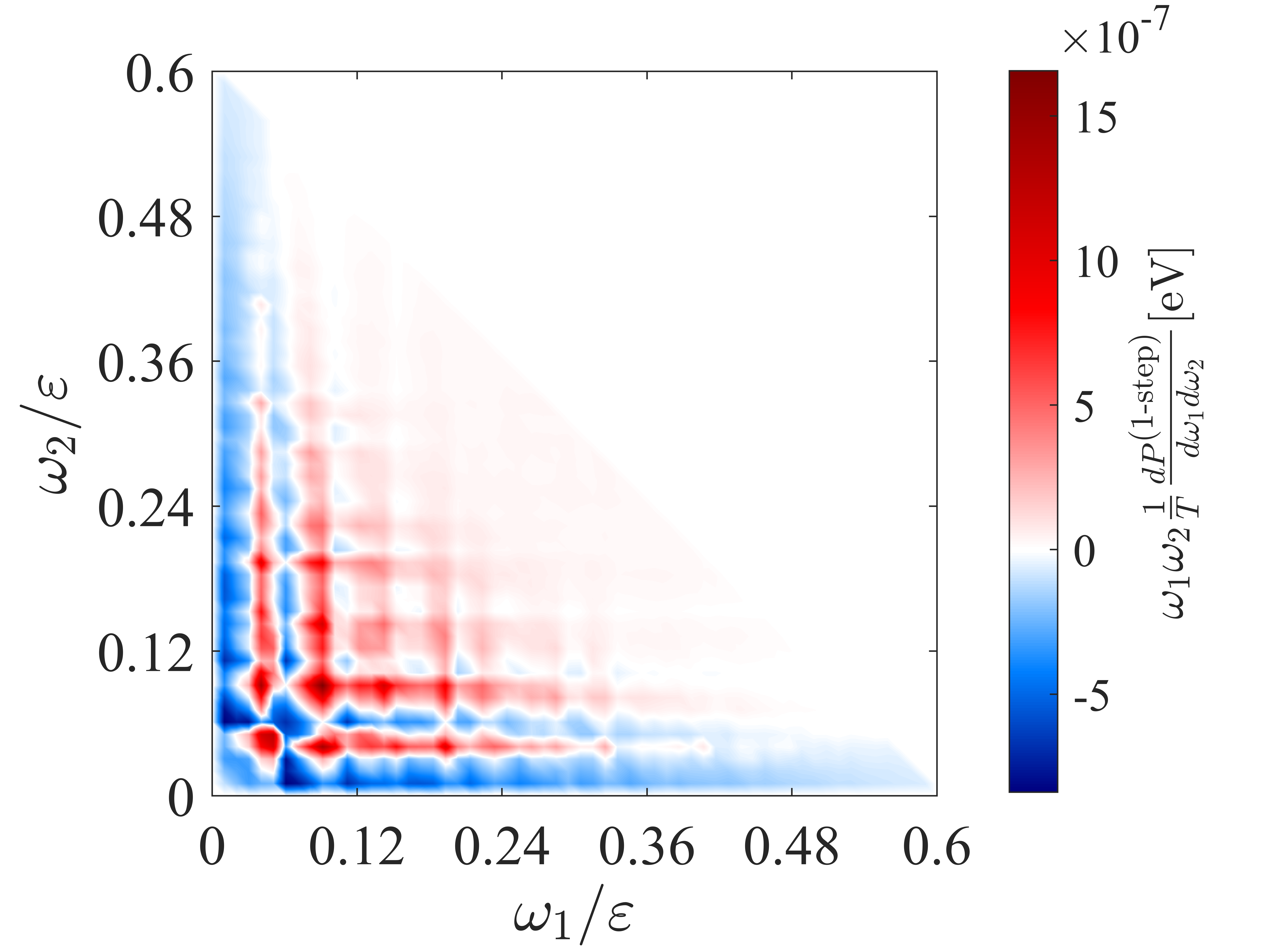}

\caption{The differential emission probability of two photons with energy $\omega_{1}$
and $\omega_{2}$ divided by $T$ for the one-step contribution, for
the case mentioned in the text and as in figure (\ref{fig:Cascadeinf}).\label{fig:Interference}}
\end{figure}

\section{Discussion of results}

\begin{figure}[t]
\includegraphics[width=1\columnwidth]{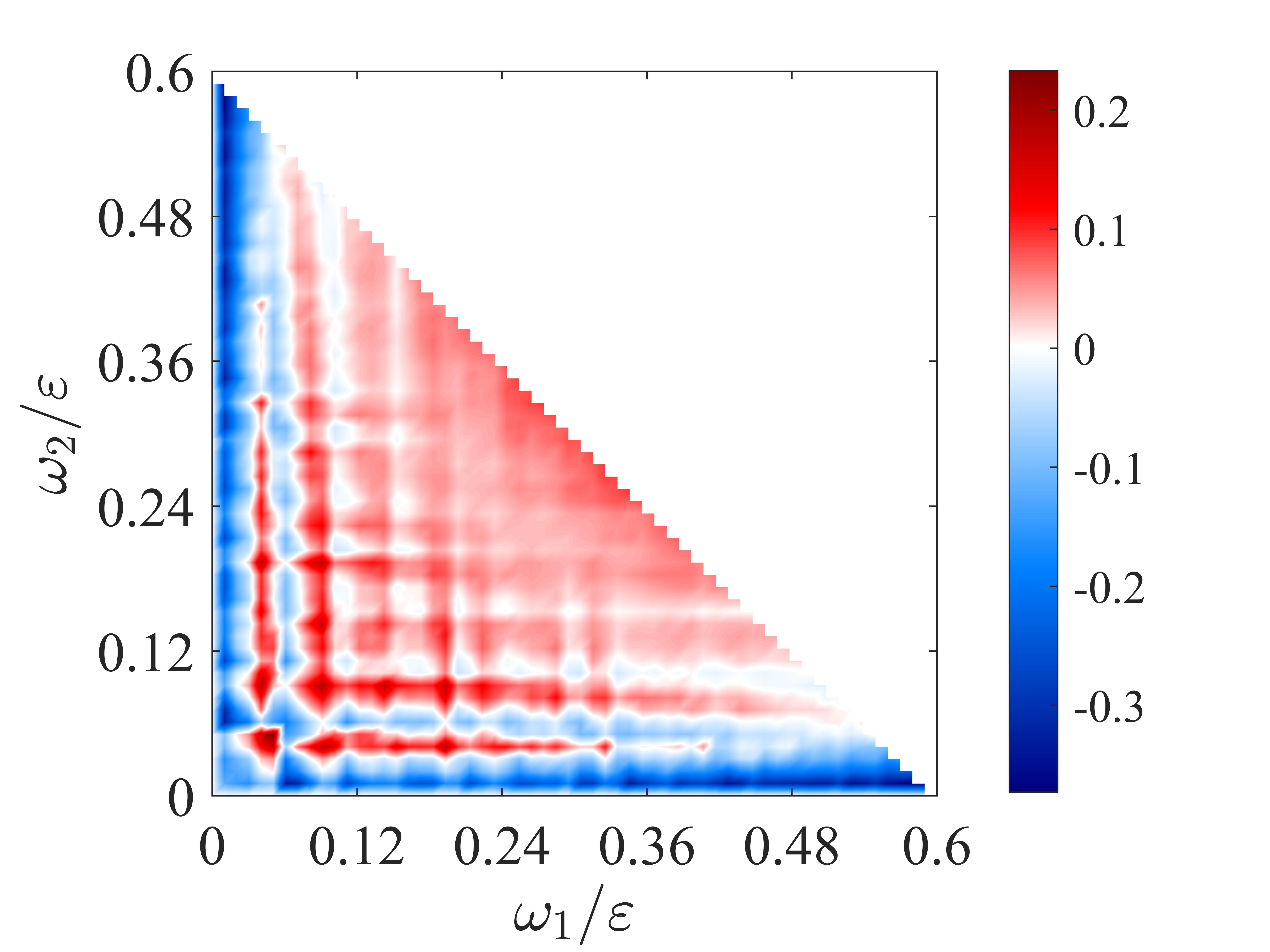}

\caption{The ratio of emission probabilities of two photons with energy $\omega_{1}$
and $\omega_{2}$ for the one-step contribution to the cascade contribution
when $T=20\mu$m, for the case mentioned in the text and as in figure
(\ref{fig:Cascadeinf}). This ratio therefore scales as $1/T$.\label{fig:Ratio}}
\end{figure}

In the figures in this paper we show the calculations made for a 180
GeV electron in the Doyle-Turner potential \cite{Baier1998,Doyle,avakian1982,Moller1995403}
for the (110) planes in Silicon and for the state $n=25$. This is
a quite low lying state which for electrons will have a high radiation
power \cite{Wistisen2019}. Electrons were chosen for this reason
as it is not as numerically heavy when the quantum numbers are relatively
small, as opposed to the positron case, which would require large
quantum numbers to obtain an appreciable value of the quantum non-linearity
parameter $\chi$, which means that quantum effects such as spin and
recoil are important in the emission process. To compare with an experiment
one should average over the distribution of the initial states which
depends on the particle beam angular mean and divergence. In (\ref{eq:Fullprob-1})
the integrals over $\varphi$ and $\theta$ are carried out numerically
over the intervals $0<\varphi<2\pi$ and $0<\theta<\frac{1.5}{\gamma}\times(1+\xi)$,
and therefore includes nearly all emitted radiation. From the result
of Eq. (\ref{eq:Fullprob-1}) we see that the part scaling with $T^{2}$
is the cascade, obtained by simple multiplication of probabilities,
and will dominate unless the crystal is very thin, due to the remaining
terms being proportional to $T$. Therefore, if one made a Monte Carlo
approach using the single photon emission rate using the quantum numbers
of the current state, instead of using the constant field approximation
with the current value of the field, one would obtain the dominant
(cascade) contribution, which will be accurate also when the constant
field approximation is no longer valid. In figure (\ref{fig:Cascadeinf})
we show the result from the cascade process. In figure (\ref{fig:Interference})
we show the one-step terms and finally in figure (\ref{fig:Ratio})
we show the ratio of these one-step terms to the cascade terms for
$T=20\text{\textmu m}$. From this figure we see that the one-step
terms can become significant compared to the cascade terms for short
crystals. This ratio scales as $1/T$. Therefore one needs a thin
crystal for the one-step contribution to be significant, so thin that
the probability to emit more than 1 photon becomes small. One may
rightfully ask based on these figures, if one picks a very small value
of $T$, the total probability could seemingly become negative, however
the results shown are only valid when $T\gg l_{f}\sim0.8$ \textmu m
as estimated earlier. For the $180$ GeV case calculated here, the
probability to emit a photon with energy above $1$ GeV from a 20
\textmu m crystal is roughly $7\%$ and therefore the probability
corresponding to the cascade for two-photon emission above this photon
energy is $0.25\%$, and as can be seen in figure (\ref{fig:Ratio})
the spectrum in the region where the radiation is most abundant, the
ratio is around $\pm20\%$. This number serves as an upper limit to
the size of the effect, because under experimental conditions one
would obtain the average from a population of many different levels
with different quantum number $n$, and this averaging would likely
reduce the size of the effect. If we assume the size of the effect
to be this upper limit, one would need enough events such that one
would have enough statistics to see an effect of such a size from
only $0.25\%$ of the events. If this setup was realized by adding
a calorimeter to a setup as the one used in \cite{wistisen2018experimental}
we can estimate the number of particles required to see this. 
Making a histogram of 20 bins in each direction of $\omega_{1}$ and
$\omega_{2}$ and assuming 100 counts on average in each bin, one
would need roughly $3.2\times10^{8}$ electrons and assuming an electron
rate of $10^{4}/\text{min}$ this translates into roughly 22 days
of measuring time. This would therefore be a challenging experiment
and having in mind that there would likely also be systematic uncertainties,
the realistic outcome of such an experiment would be to put a constraint
on the size of such one-step terms, rather than their direct observation.

\section{Conclusion}

In conclusion, we have shown how to accurately calculate the two photon
emission rate for a high energy electron (or positron) channeled in
a crystal. This calculation shows that the full probability contains
what is known as the cascade, which could have been obtained multiplying
probabilities of single photon emissions, as well as additional interference
terms, called the one-step contribution. The one-step contribution
scales only linearly with the crystal length, and therefore one needs
a thin crystal to see the effect of these terms. We have calculated
the size of all contributions to the emission probability for 180
GeV electrons in Silicon and found that with a long measuring time,
the one-step contribution could possibly be seen. Since these effects
are however small, we also see how to solve the problem of quantum
radiation reaction, under general circumstances, in a crystal, by
using the single photon emission rate in consecutive emissions, corresponding
to the particle's current state.

\newpage{}

\section{Acknowledgments}

The author gratefully acknowledges useful discussions with Antonino
Di Piazza and Karen Z. Hatsagortsyan. This work was partially supported
by a research grant (VKR023371) from VILLUM FONDEN and later by the
Alexander von Humboldt-Stiftung. In addition the author acknowledges
the support of NVIDIA Corporation with the donation of the Titan V
GPU used for this research.

\begin{widetext}

\section*{Appendix A}

The general (unnormalized) solution to the Dirac equation with potential
energy $V(\boldsymbol{r})=-e\varphi(\boldsymbol{r})$ can be written
as

\begin{equation}
\psi(\boldsymbol{r},t)=e^{-i\varepsilon t}\left(\begin{array}{c}
\phi(\boldsymbol{r})\\
\chi(\boldsymbol{r})
\end{array}\right)
\end{equation}
The Dirac equation then becomes

\begin{equation}
\left(\varepsilon+e\varphi-m\right)\phi(\boldsymbol{r})=\boldsymbol{\sigma}\cdot\hat{\boldsymbol{p}}\chi(\boldsymbol{r})\label{eq:Dirac1}
\end{equation}

\begin{equation}
\left(\varepsilon+e\varphi+m\right)\chi(\boldsymbol{r})=\boldsymbol{\sigma}\cdot\hat{\boldsymbol{p}}\phi(\boldsymbol{r})\label{eq:Dirac2}
\end{equation}
The electron solution is then

\begin{equation}
\psi^{-}(\boldsymbol{r},t)=e^{-i\varepsilon t}\left(\begin{array}{c}
\phi(\boldsymbol{r})\\
\frac{\boldsymbol{\sigma}\cdot\hat{\boldsymbol{p}}}{\varepsilon+e\varphi(\boldsymbol{r})+m}\phi(\boldsymbol{r})
\end{array}\right)
\end{equation}
We then obtained an equation for $\phi(\boldsymbol{r})$ by isolating
$\chi(\boldsymbol{r})$ in Eq. (\ref{eq:Dirac2}) and inserting in
Eq. (\ref{eq:Dirac1}). This solution has the property that it is
well defined when $\varepsilon=m$. Another solution can be found
by isolating $\chi(\boldsymbol{r})$ in Eq. (\ref{eq:Dirac1}) and
inserting in (\ref{eq:Dirac2}). However this solution is not well
defined when $\varepsilon=m$ and therefore one must use the negative
energy solution $\varepsilon=-\sqrt{p^{2}+m^{2}}=-\varepsilon_{p}$
therefore we have
\begin{equation}
\psi(\boldsymbol{r},t)=e^{i\varepsilon_{p}t}\left(\begin{array}{c}
-\frac{\boldsymbol{\sigma}\cdot\hat{\boldsymbol{p}}}{\varepsilon_{p}-e\varphi(\boldsymbol{r})+m}\chi(\boldsymbol{r})\\
\chi(\boldsymbol{r})
\end{array}\right)
\end{equation}
where $\varepsilon_{p}$ is the positive energy of the positron. The
equation for $\chi$ we can now be obtained by using

\begin{equation}
\left(\varepsilon+e\varphi+m\right)\chi(\boldsymbol{r})=\boldsymbol{\sigma}\cdot\hat{\boldsymbol{p}}\frac{1}{\varepsilon+e\varphi-m}\boldsymbol{\sigma}\cdot\hat{\boldsymbol{p}}\chi(\boldsymbol{r}),
\end{equation}
which is equivalent with

\begin{equation}
\left(\varepsilon_{p}-e\varphi-m\right)\chi(\boldsymbol{r})=\boldsymbol{\sigma}\cdot\hat{\boldsymbol{p}}\frac{1}{\varepsilon_{p}-e\varphi+m}\boldsymbol{\sigma}\cdot\hat{\boldsymbol{p}}\chi(\boldsymbol{r}).
\end{equation}
This is the same equation as the one we obtained for $\phi(\boldsymbol{r})$,
except with the sign of $e$ changed such that, after making the same
approximations as we did in \cite{PhysRevA.98.022131}:

\begin{equation}
\left[\hat{\boldsymbol{p}}^{2}+2\varepsilon_{p}e\varphi(\boldsymbol{r})-(\varepsilon_{p}^{2}-m^{2})\right]\chi(\boldsymbol{r})=0
\end{equation}
We therefore make the ansatz in line with the usual approach (the
sign on the momenta is changed):

\begin{equation}
\chi(\boldsymbol{r})=\boldsymbol{s}I^{+}(y)e^{-i(p_{x}x+p_{z}z)}
\end{equation}

\begin{equation}
I^{+}(y)=e^{-ik_{B}y}\sum_{j}c_{j}e^{-ijk_{0}y}.
\end{equation}
Then

\begin{equation}
\psi(\boldsymbol{r},t)=e^{i(-p_{x}x-p_{z}z+\varepsilon_{p}t)}\left(\begin{array}{c}
\frac{\boldsymbol{\sigma}\cdot\boldsymbol{p}}{\varepsilon_{p}+m}\boldsymbol{s}I^{+}(y)\\
\boldsymbol{s}I^{+}(y)
\end{array}\right),
\end{equation}
with $\boldsymbol{p}=(p_{x}+q\varphi(\boldsymbol{r}),i\frac{d}{dy},p_{z})$,
inserting $I^{+}(y)$, this becomes

\begin{equation}
\psi(\boldsymbol{r},t)=e^{i(-p_{x}x-p_{z}z+\varepsilon_{p}t)}\sum_{j}c_{j}e^{-i(k_{B}+jk_{0})y}\left(\begin{array}{c}
\frac{\boldsymbol{\sigma}\cdot\boldsymbol{p}}{\varepsilon_{p}+m}\boldsymbol{s}\\
\boldsymbol{s}
\end{array}\right),
\end{equation}
with $\boldsymbol{p}=(p_{x}+E_{n}-\frac{(jk_{0}+k_{B})^{2}}{2\varepsilon},jk_{0}+k_{B},p_{z})$.

\section*{Appendix B}

The electron state can be written as (putting back in the volume factor)

\begin{equation}
\psi_{p,s}(x)=\frac{1}{\sqrt{2\varepsilon V}}e^{-i\varepsilon_{n}t}e^{i(p_{x}x+k_{B}y+p_{z}z)}\sum_{j}c_{j}\boldsymbol{S}_{j}e^{ijk_{0}y},\label{eq:electronwavefunc-1}
\end{equation}
where

\begin{equation}
\boldsymbol{S}_{j}=\sqrt{\varepsilon+m}\left(\begin{array}{c}
\boldsymbol{s}\\
\frac{\boldsymbol{\sigma}\cdot\boldsymbol{p}_{j}}{\varepsilon+m}\boldsymbol{s}
\end{array}\right),\label{eq:bigS}
\end{equation}
where $\boldsymbol{p}_{j}=(p_{x}+E_{n}-\frac{(jk_{0}+k_{B})^{2}}{2\varepsilon},jk_{0}+k_{B},p_{z})$
and then

\begin{equation}
\int\psi_{p'}^{\dagger}\psi_{p}dV=\frac{1}{2V\sqrt{\varepsilon'\varepsilon}}(2\pi)^{3}\delta(p_{x}-p_{x}')\delta(p_{z}-p_{z}')\sum_{j,j'}c_{j}(p)c_{j'}^{*}(p')\boldsymbol{S}_{j'}^{'\dagger}\boldsymbol{S}_{j}\delta(k_{B}-k_{B}'+(j-j')k_{0}).
\end{equation}
Explicitly we have that $c_{j}=c_{j}(p_{x},k_{B},p_{z},n)$. Now since
both $k_{B}$ and $k_{B}'$ obey that $0\leq k_{B}<k_{0}$ we have
that $-k_{0}<k_{B}-k_{B}'<k_{0}$ and therefore $k_{B}-k_{B}'$ can
never be an integer value of $k_{0}$ unless $k_{B}-k_{B}'=0$, and
therefore we can write

\begin{equation}
\delta(k_{B}-k_{B}'+(j-j')k_{0})=\delta(k_{B}-k_{B}')\delta_{j,j'}
\end{equation}

\begin{equation}
\int\psi_{p'}^{\dagger}\psi_{p}dV=\frac{1}{2V\sqrt{\varepsilon'\varepsilon}}(2\pi)^{3}\delta(p_{x}-p_{x}')\delta(p_{z}-p_{z}')\delta(k_{B}-k_{B}')\sum_{j}c_{j}(p_{x},k_{B},p_{z},n)c_{j}^{*}(p_{x},k_{B},p_{z},n')\boldsymbol{S}_{j}^{'\dagger}\boldsymbol{S}_{j}
\end{equation}
However the vector $\boldsymbol{c}$ is a normalized ($|\boldsymbol{c}|=1$),
eigenvector of a hermitian matrix and the vectors corresponding to
$n$ and $n'$ have different eigenvalues of this matrix, and are
therefore orthogonal, so

\begin{equation}
\int\psi_{p'}^{\dagger}\psi_{p}dV=\frac{1}{2V\sqrt{\varepsilon'\varepsilon}}(2\pi)^{3}\delta(p_{x}-p_{x}')\delta(p_{z}-p_{z}')\delta(k_{B}-k_{B}')\delta_{n,n'}\sum_{j}|c_{j}|^{2}\boldsymbol{S}_{j}^{'\dagger}\boldsymbol{S}_{j}.
\end{equation}
Now consider

\begin{equation}
\boldsymbol{S}_{j}^{'\dagger}\boldsymbol{S}_{j}=(\varepsilon+m)\left(\boldsymbol{s}^{'\dagger}\boldsymbol{s}+\boldsymbol{s}^{'\dagger}\frac{\boldsymbol{\sigma}\cdot\boldsymbol{p}_{j}}{\varepsilon+m}\frac{\boldsymbol{\sigma}\cdot\boldsymbol{p}_{j}}{\varepsilon+m}\boldsymbol{s}\right)=\boldsymbol{s}^{'\dagger}\boldsymbol{s}\left[(\varepsilon+m)+\frac{\boldsymbol{p}_{j}^{2}}{\varepsilon+m}\right],
\end{equation}
and therefore
\begin{equation}
\sum_{j}|c_{j}|^{2}\boldsymbol{S}_{j}^{'\dagger}\boldsymbol{S}_{j}\simeq2\varepsilon\delta_{s',s}.
\end{equation}
There $\simeq$ refers only to the normalization. The states are exactly
orthogonal, but in the normalization we neglect corrections which
are suppressed by at least $\xi/\gamma$ compared to leading order.
So finally

\begin{equation}
\int\psi_{p'}^{\dagger}\psi_{p}dV=\frac{(2\pi)^{3}}{V}\delta(p_{x}-p_{x}')\delta(p_{z}-p_{z}')\delta(k_{B}-k_{B}')\delta_{n,n'}\delta_{s',s}
\end{equation}

\section*{Appendix C}

Even though we consider the radiation from electrons, the propagator
contains terms from the positron $\psi_{p,n,s}^{+}(\boldsymbol{x}_{2})\bar{\psi}_{p,n,s}^{+}(\boldsymbol{x}_{1})$.
Therefore we will need to calculate

\begin{equation}
\mathcal{M}_{i\rightarrow v}^{+}=e\sqrt{\frac{4\pi}{2\omega}}\frac{1}{2\sqrt{\varepsilon_{v}\varepsilon_{i}}}\sum_{j,l}c_{l,v}^{*}c_{j,i}\bar{\boldsymbol{S}}_{l,v}^{+}\slashed{\epsilon}^{*}\boldsymbol{S}_{j,i}^{-}
\end{equation}
and $\mathcal{M}_{v\rightarrow f}^{+}$ so we need

\begin{flalign}
 & -\frac{1}{\sqrt{\varepsilon_{i}+m}\sqrt{\varepsilon_{v}+m}}\bar{\boldsymbol{S}}_{v}^{+}\slashed{\epsilon}^{*}\boldsymbol{S}_{i}^{-}\nonumber \\
 & =\left[\left(\begin{array}{cc}
\boldsymbol{s}_{v}^{T}\frac{\boldsymbol{\sigma}\cdot\boldsymbol{p}_{v}}{\varepsilon_{v}+m} & \boldsymbol{s}_{v}^{T}\end{array}\right)\left(\begin{array}{cc}
0 & \boldsymbol{\sigma}\cdot\boldsymbol{\epsilon}^{*}\\
\boldsymbol{\sigma}\cdot\boldsymbol{\epsilon}^{*} & 0
\end{array}\right)\left(\begin{array}{c}
\boldsymbol{s}_{i}\\
\frac{\boldsymbol{\sigma}\cdot\boldsymbol{p}_{i}}{\varepsilon_{i}+m}\boldsymbol{s}_{i}
\end{array}\right)\right]\nonumber \\
 & =\left[\left(\begin{array}{cc}
\boldsymbol{s}_{v}^{T}\frac{\boldsymbol{\sigma}\cdot\boldsymbol{p}_{v}}{\varepsilon_{v}+m} & \boldsymbol{s}_{v}^{T}\end{array}\right)\left(\begin{array}{c}
\boldsymbol{\sigma}\cdot\boldsymbol{\epsilon}^{*}\frac{\boldsymbol{\sigma}\cdot\boldsymbol{p}_{i}}{\varepsilon_{i}+m}\boldsymbol{s}_{i}\\
\boldsymbol{\sigma}\cdot\boldsymbol{\epsilon}^{*}\boldsymbol{s}_{i}
\end{array}\right)\right]\nonumber \\
 & =\boldsymbol{s}_{v}^{T}\left[\frac{\boldsymbol{\sigma}\cdot\boldsymbol{p}_{v}}{\varepsilon_{v}+m}\boldsymbol{\sigma}\cdot\boldsymbol{\epsilon}^{*}\frac{\boldsymbol{\sigma}\cdot\boldsymbol{p}_{i}}{\varepsilon_{i}+m}+\boldsymbol{\sigma}\cdot\boldsymbol{\epsilon}^{*}\right]\boldsymbol{s}_{i}\nonumber \\
 & =\boldsymbol{s}_{v}^{T}\left[\frac{1}{(\varepsilon_{i}+m)(\varepsilon_{v}+m)}\left(\boldsymbol{\sigma}\cdot\boldsymbol{p}_{v}\right)\left(\boldsymbol{\sigma}\cdot\boldsymbol{\epsilon}^{*}\right)\left(\boldsymbol{\sigma}\cdot\boldsymbol{p}_{i}\right)+\boldsymbol{\sigma}\cdot\boldsymbol{\epsilon}^{*}\right]\boldsymbol{s}_{i}\nonumber \\
 & =\boldsymbol{s}_{v}^{T}\left[\frac{1}{(\varepsilon_{i}+m)(\varepsilon_{v}+m)}\left(\boldsymbol{\sigma}\cdot\boldsymbol{p}_{v}\right)\left(\boldsymbol{\epsilon}^{*}\cdot\boldsymbol{p}_{i}+i\boldsymbol{\sigma}\cdot\left[\boldsymbol{\epsilon}^{*}\times\boldsymbol{p}_{i}\right]\right)+\boldsymbol{\sigma}\cdot\boldsymbol{\epsilon}^{*}\right]\boldsymbol{s}_{i}\nonumber \\
 & =\boldsymbol{s}_{v}^{T}\left[\frac{1}{(\varepsilon_{i}+m)(\varepsilon_{v}+m)}\left\{ \left(\boldsymbol{\sigma}\cdot\boldsymbol{p}_{v}\right)\left(\boldsymbol{\epsilon}^{*}\cdot\boldsymbol{p}_{i}\right)+i\left(\boldsymbol{\sigma}\cdot\boldsymbol{p}_{v}\right)\boldsymbol{\sigma}\cdot\left[\boldsymbol{\epsilon}^{*}\times\boldsymbol{p}_{i}\right]\right\} +\boldsymbol{\sigma}\cdot\boldsymbol{\epsilon}^{*}\right]\boldsymbol{s}_{i}\nonumber \\
 & =\boldsymbol{s}_{v}^{T}\left[\frac{1}{(\varepsilon_{i}+m)(\varepsilon_{v}+m)}\left\{ \left(\boldsymbol{\sigma}\cdot\boldsymbol{p}_{v}\right)\left(\boldsymbol{\epsilon}^{*}\cdot\boldsymbol{p}_{i}\right)+i\left[\boldsymbol{p}_{v}\cdot\left(\boldsymbol{\epsilon}^{*}\times\boldsymbol{p}_{i}\right)+i\boldsymbol{\sigma}\cdot\left(\boldsymbol{p}_{v}\times\left(\boldsymbol{\epsilon}^{*}\times\boldsymbol{p}_{i}\right)\right)\right]\right\} +\boldsymbol{\sigma}\cdot\boldsymbol{\epsilon}^{*}\right]\boldsymbol{s}_{i}\nonumber \\
 & =\boldsymbol{s}_{v}^{T}\left[\frac{1}{(\varepsilon_{i}+m)(\varepsilon_{v}+m)}\left\{ \left(\boldsymbol{\sigma}\cdot\boldsymbol{p}_{v}\right)\left(\boldsymbol{\epsilon}^{*}\cdot\boldsymbol{p}_{i}\right)+i\boldsymbol{p}_{v}\cdot\left(\boldsymbol{\epsilon}^{*}\times\boldsymbol{p}_{i}\right)-\boldsymbol{\sigma}\cdot\left(\boldsymbol{p}_{v}\times\left(\boldsymbol{\epsilon}^{*}\times\boldsymbol{p}_{i}\right)\right)\right\} +\boldsymbol{\sigma}\cdot\boldsymbol{\epsilon}^{*}\right]\boldsymbol{s}_{i}\nonumber \\
 & =\boldsymbol{s}_{v}^{T}\left[iC+\boldsymbol{\sigma}\cdot\boldsymbol{D}\right]\boldsymbol{s}_{i}
\end{flalign}
Then

\begin{equation}
C=\frac{\boldsymbol{p}_{v}\cdot\left(\boldsymbol{\epsilon}^{*}\times\boldsymbol{p}_{i}\right)}{(\varepsilon_{i}+m)(\varepsilon_{v}+m)}=\frac{\boldsymbol{\epsilon}^{*}\cdot\left(\boldsymbol{p}_{i}\times\boldsymbol{p}_{v}\right)}{(\varepsilon_{i}+m)(\varepsilon_{v}+m)},
\end{equation}

\begin{equation}
\boldsymbol{D}=\frac{\left(\boldsymbol{\epsilon}^{*}\cdot\boldsymbol{p}_{i}\right)\boldsymbol{p}_{v}-\boldsymbol{p}_{v}\times\left(\boldsymbol{\epsilon}^{*}\times\boldsymbol{p}_{i}\right)}{(\varepsilon_{i}+m)(\varepsilon_{v}+m)}+\boldsymbol{\epsilon}^{*}.
\end{equation}
Here we may use that $\boldsymbol{p}_{v}\times\left(\boldsymbol{\epsilon}^{*}\times\boldsymbol{p}_{i}\right)=\boldsymbol{\epsilon}^{*}(\boldsymbol{p}_{i}\cdot\boldsymbol{p}_{v})-\boldsymbol{p}_{i}(\boldsymbol{p}_{v}\cdot\boldsymbol{\epsilon}^{*})$
and so

\begin{equation}
\boldsymbol{D}=\frac{\left(\boldsymbol{\epsilon}^{*}\cdot\boldsymbol{p}_{i}\right)\boldsymbol{p}_{v}+\boldsymbol{p}_{i}(\boldsymbol{p}_{v}\cdot\boldsymbol{\epsilon}^{*})}{(\varepsilon_{i}+m)(\varepsilon_{v}+m)}+\boldsymbol{\epsilon}^{*}\left(1-\frac{\boldsymbol{p}_{i}\cdot\boldsymbol{p}_{v}}{(\varepsilon_{i}+m)(\varepsilon_{v}+m)}\right).
\end{equation}
Now consider the other part for $\mathcal{M}_{v\rightarrow f}^{+}(\boldsymbol{k},\epsilon)$

\begin{flalign}
-\frac{1}{\sqrt{\varepsilon_{f}+m}\sqrt{\varepsilon_{v}+m}}\bar{\boldsymbol{S}}_{f}^{-}\slashed{\epsilon}^{*}\boldsymbol{S}_{v}^{+} & =\left[\left(\begin{array}{cc}
\boldsymbol{s}_{f}^{T} & \boldsymbol{s}_{f}^{T}\frac{\boldsymbol{\sigma}\cdot\boldsymbol{p}_{f}}{\varepsilon_{f}+m}\end{array}\right)\left(\begin{array}{cc}
0 & \boldsymbol{\sigma}\cdot\boldsymbol{\epsilon}^{*}\\
\boldsymbol{\sigma}\cdot\boldsymbol{\epsilon}^{*} & 0
\end{array}\right)\left(\begin{array}{c}
\frac{\boldsymbol{\sigma}\cdot\boldsymbol{p}_{v}}{\varepsilon_{v}+m}\boldsymbol{s}_{v}\\
\boldsymbol{s}_{v}
\end{array}\right)\right]\nonumber \\
 & =\left[\left(\begin{array}{cc}
\boldsymbol{s}_{f}^{T} & \boldsymbol{s}_{f}^{T}\frac{\boldsymbol{\sigma}\cdot\boldsymbol{p}_{f}}{\varepsilon_{f}+m}\end{array}\right)\left(\begin{array}{c}
\boldsymbol{\sigma}\cdot\boldsymbol{\epsilon}^{*}\boldsymbol{s}_{v}\\
\boldsymbol{\sigma}\cdot\boldsymbol{\epsilon}^{*}\frac{\boldsymbol{\sigma}\cdot\boldsymbol{p}_{v}}{\varepsilon_{v}+m}\boldsymbol{s}_{v}
\end{array}\right)\right]\nonumber \\
 & =\boldsymbol{s}_{f}^{T}\left[\boldsymbol{\sigma}\cdot\boldsymbol{\epsilon}^{*}+\frac{\boldsymbol{\sigma}\cdot\boldsymbol{p}_{f}}{\varepsilon_{f}+m}\boldsymbol{\sigma}\cdot\boldsymbol{\epsilon}^{*}\frac{\boldsymbol{\sigma}\cdot\boldsymbol{p}_{v}}{\varepsilon_{v}+m}\right]\boldsymbol{s}_{v}.
\end{flalign}
This is the same as before except with $i\rightarrow v$ and $v\rightarrow f$
. And now we want the quantity

\begin{align}
 & \sum_{j,l}c_{l,v}^{*}c_{j,i}\bar{\boldsymbol{S}}_{l,v}^{+}\slashed{\epsilon}^{*}\boldsymbol{S}_{j,i}^{-}e^{i(k_{B,v}+k_{B,i}-k_{y})y}e^{i(j+l)k_{0}y}\nonumber \\
 & =2\pi\delta(k_{B,v}+k_{B,i}-k_{y}-n_{B,1}^{+}k_{0})\sum_{j}c_{-(n_{B,1}^{+}+j),v}^{*}c_{j,i}\bar{\boldsymbol{S}}_{-(n_{B,1}^{+}+j),v}^{+}\slashed{\epsilon}^{*}\boldsymbol{S}_{j,i}^{-}
\end{align}
where now $n_{B,1}^{+}$ is chosen such that $k_{B,v}=k_{y}-k_{B,i}+n_{B,1}^{+}k_{0}$
is in the FBZ. Note that $-k_{B,v}=k_{B,i}-k_{y}-n_{B,1}^{+}k_{0}$
for which we already have the solution, called $k_{B,v}^{-}=k_{B,i}-k_{y}-n_{B,1}^{-}k_{0}$,
and therefore

\begin{equation}
k_{B,v}=-k_{B,v}^{-}+k_{0}=-k_{B,i}+k_{y}+(n_{B,1}^{-}+1)k_{0}
\end{equation}
therefore $n_{B,1}^{+}=n_{B,1}^{-}+1$. For the $\mathcal{M}_{v\rightarrow f}^{+}$
term one obtains that $k_{B,f}=-k_{B,v}-k_{y}-n_{B,2}^{+}k_{0}$ and
for this term one has that $n_{B,2}^{+}=n_{B,2}^{-}-1$, in terms
of the $n_{B,2}^{-}$ value for the corresponding electron term in
the propagator. And that the $l$ index is given by $l=n_{B,2}^{+}-j$.

\section*{Appendix D}

We need to consider $|\sum_{s_{v}}\mathcal{M}_{2}\mathcal{M}_{1}|^{2}$,
in particular we would like to show that $\text{Re}\left(\left[\mathcal{M}_{2,\uparrow}\mathcal{M}_{1,\uparrow}\right]\left[\mathcal{M}_{2,\downarrow}\mathcal{M}_{1,\downarrow}\right]^{\dagger}\right)$
is 0, where the arrows denote the spin state of the virtual particle.
This we may rearrange and consider therefore the product $\mathcal{M}_{2,\downarrow}^{\dagger}\mathcal{M}_{2,\uparrow}$.
Now we may use that $\mathcal{M}$ can be written as 

\begin{align}
\mathcal{M}_{2} & =e\sqrt{\frac{4\pi}{2\omega}}\frac{1}{2\sqrt{\varepsilon_{f}\varepsilon_{i}}}\sum_{j}c_{n_{B,2}+j,f}^{*}c_{j,v}\bar{\boldsymbol{S}}_{n_{B,2}+j,f}\slashed{\epsilon}^{*}\boldsymbol{S}_{j,v}\nonumber \\
 & =-e\sqrt{\frac{4\pi}{2\omega}}\frac{1}{2\sqrt{\varepsilon_{f}\varepsilon_{i}}}\sum_{j}c_{n_{B,2}+j,f}^{*}c_{j,v}\boldsymbol{s}_{f}^{\dagger}\left[\boldsymbol{\epsilon}^{*}\cdot\boldsymbol{A}_{n_{B,2}+j,j}+i\boldsymbol{B}_{n_{B,2}+j,j}\cdot\boldsymbol{\sigma}\right]\boldsymbol{s}_{v}.
\end{align}
Now for simplicity we define 
\begin{align}
\tilde{\boldsymbol{A}}=-e\sqrt{\frac{4\pi}{2\omega}}\frac{1}{2\sqrt{\varepsilon_{f}\varepsilon_{i}}}\sum_{j}c_{n_{B,2}+j,f}^{*}c_{j,v}\boldsymbol{A}_{n_{B,2}+j,j},
\end{align}
\begin{align}
\tilde{\boldsymbol{B}}=-e\sqrt{\frac{4\pi}{2\omega}}\frac{1}{2\sqrt{\varepsilon_{f}\varepsilon_{i}}}\sum_{j}c_{n_{B,2}+j,f}^{*}c_{j,v}\boldsymbol{B}_{n_{B,2}+j,j},
\end{align}
and then we have that
\begin{align}
\mathcal{M}_{2} & =\boldsymbol{s}_{f}^{\dagger}\left[\boldsymbol{\epsilon}^{*}\cdot\tilde{\boldsymbol{A}}+i\tilde{\boldsymbol{B}}\cdot\boldsymbol{\sigma}\right]\boldsymbol{s}_{v}.
\end{align}
Therefore

\begin{align}
\mathcal{M}_{2,\downarrow}^{\dagger}\mathcal{M}_{2,\uparrow} & =\boldsymbol{s}_{\downarrow}^{\dagger}\left[\boldsymbol{\epsilon}^{*}\cdot\tilde{\boldsymbol{A}}+i\tilde{\boldsymbol{B}}\cdot\boldsymbol{\sigma}\right]^{\dagger}\boldsymbol{s}_{f}\boldsymbol{s}_{f}^{\dagger}\left[\boldsymbol{\epsilon}^{*}\cdot\tilde{\boldsymbol{A}}+i\tilde{\boldsymbol{B}}\cdot\boldsymbol{\sigma}\right]\boldsymbol{s}_{\uparrow}
\end{align}
We assume that $\boldsymbol{\epsilon}^{*}=\boldsymbol{\epsilon}$,
which is possible if we choose linear polarization as our basis, and
we will perform the summation of final spins and therefore $\boldsymbol{s}_{f}\boldsymbol{s}_{f}^{\dagger}$
is the identity

\begin{align}
\mathcal{M}_{2,\downarrow}^{\dagger}\mathcal{M}_{2,\uparrow} & =\boldsymbol{s}_{\downarrow}^{\dagger}\left[\boldsymbol{\epsilon}\cdot\tilde{\boldsymbol{A}}-i\tilde{\boldsymbol{B}}\cdot\boldsymbol{\sigma}\right]\left[\boldsymbol{\epsilon}\cdot\tilde{\boldsymbol{A}}+i\tilde{\boldsymbol{B}}\cdot\boldsymbol{\sigma}\right]\boldsymbol{s}_{\uparrow}\nonumber \\
 & =\boldsymbol{s}_{\downarrow}^{\dagger}\left[(\boldsymbol{\epsilon}\cdot\tilde{\boldsymbol{A}})^{2}+\tilde{\boldsymbol{B}}^{2}\right]\boldsymbol{s}_{\uparrow}\nonumber \\
 & =0
\end{align}
where we used that $\tilde{\boldsymbol{B}}$ is a real vector. For
the other term, $\mathcal{M}_{1,\uparrow}\mathcal{M}_{1,\downarrow}^{\dagger}$
the same can be done, and here the argument hinges upon summation
over initial spins, therefore, if either a summation is carried out
over initial or final spins, the spin interference terms will be 0.

\end{widetext}

\bibliography{biblio}

\end{document}